\documentclass[paper,11pt]{JHEP3}

\usepackage{amssymb,amsfonts, mathrsfs ,amsthm,latexsym,graphicx,verbatim,epsfig}

\usepackage[fleqn]{amsmath}

\title{Vacuum stability, string density of states and the Riemann zeta function}

\author{
     Carlo Angelantonj$^{a}$,  Matteo Cardella$^{b}$, Shmuel Elitzur$^{c}$ and  Eliezer Rabinovici$^{c}$\\
$^{a}$ Dipartimento di Fisica Teorica, Universit\`a di Torino, and INFN Sezione di Torino\\
Via P. Giuria 1, 10125 Torino, Italy\\
$^{b}$Instituut voor Theoretische Fysica, Valckenierstraat 65,
1018XE Amsterdam, The Netherlands.\\
 $^{c}$  Racah Institute of Physics,
    Hebrew University
    Jerusalem 91904,
    Israel.}

\abstract{
We study the distribution of graded degrees of freedom in classically stable oriented closed string vacua and use the Rankin-Selberg transform to link it to the finite one-loop vacuum energy. In particular, we find that the spectrum of physical excitations not only must enjoy {\it asymptotic supersymmetry} but actually, at very large mass, bosonic and fermionic states must follow a universal oscillating pattern, whose frequencies are related to the zeros of the Riemann $\zeta$-function. Moreover, the convergence rate of the overall number of the graded degrees of freedom to the value of the vacuum energy is determined by the Riemann hypothesis. We discuss also attempts to obtain constraints in the case of tachyon-free open-string theories.
}

\preprint{DFTT 26/2010}

\begin{document}

\section{Introduction and conclusions}
\label{intro}

In Field Theory, one usually does not pay much attention to the one-loop vacuum amplitude. This is a function of the masses of the finite number of fields of a given model, fully determined by the free spectrum that, aside from its relation to the cosmological constant, does not embody important structural information. On the other hand, strings describe infinitely many modes, and their vacuum amplitudes satisfy a number of geometric constraints, that in a wide class of models essentially determine the full perturbative spectrum. 

At one loop a closed oriented string sweeps a torus, whose Teichm\"uller parameter $\tau = \tau_1 +i\tau_2$ plays the role of the (complex) Schwinger parameter, and one is instructed to integrate over all inequivalent tori. This amounts to restricting the integration over a fundamental domain and requires that the integrand function be an automorphic function of ${\rm PSL} (2,\mathbb{Z})$, the latter condition being equivalent to impose suitable GSO projections on the full string spectrum. Although the GSO truncation selects consistent string vacua and constrains the distribution of physical and unphysical states at each mass level, it is actually cumbersome to extract information about the graded number of degrees of freedom at arbitrary mass\footnote{Here and in the whole paper, by {\it graded number of degrees of freedom} we mean the difference of bosonic and fermionic ones at a given string excitation, $d_{\rm B} (n) - d_{\rm F} (n)$.}. However, the mass distribution of bosonic and fermionic excitations plays a crucial role in determining the celebrated finiteness of string theory. It is therefore important to attempt a deeper understanding of the distribution of degrees of freedom in string theory. Although space-time supersymmetry on flat space imposes a perfect equilibrium between bosonic and fermionic degrees of freedom, this is no longer evident when supersymmetry is absent or spontaneously broken. 

Most of presently known string vacua are classically unstable when space-time supersymmetry is absent. String spectra are typically plagued by the presence of tachyonic modes that suggest the onset of an instability. Although the condensation of open-string tachyons is relatively under control \cite{Sen:2004nf,Elitzur:2000pq}, very little is known about the rolling of closed-string ones \cite{Elitzur:1991cb,Antoniadis:1991kh}. Therefore, if one insists that a non-supersymmetric vacuum ought to be predictive, at least classically, one is bound to consider only string configurations where tachyonic excitations are absent. Actually, this turns out to be rather non-trivial to achieve and requires a precise distribution of graded massive string states, since these two features are strictly related by modular invariance. 

A first attempt to study the interplay between massive states and tachyons was done in \cite{Kutasov:1990sv}. It was then shown that infra-red (IR) finiteness  of the one-loop vacuum energy of closed oriented strings actually requires the presence of space-time fermions in the spectrum, and moreover their overall number, independently of the mass, must cancel against the space-time bosons almost exactly. The authors called this property {\it asymptotic supersymmetry} and put a first constraint on the {\it overall} distribution of degrees of freedom of classically stable string configurations. 

A more refined analysis of the distribution of bosonic and fermionic excitations at all energy levels was presented in 
\cite{Dienes:1994es,Dienes:1995pm}, and led to the discovery of a hidden {\it misaligned supersymmetry} that governs the arrangement of bosonic and fermionic states. Indeed, it was shown that one cannot deform the string spectrum arbitrarily, but the pattern of massive bosons and fermions must be alternating. 

In this paper we address the same problem but from a different perspective and find a remarkable connection between the graded distribution of string degrees of freedom and the pattern of the non-trivial zeroes of the Riemann $\zeta$-function, at least for the class of classically stable vacua. To be precise, this connection builds on solid mathematical ground only in the case when unphysical tachyons are absent while, based on the analysis of some concrete models and on reasonable physical assumptions, we provide some hints also in the case when unphysical tachyons are present. The content of physical (and unphysical) states in a given string vacuum\footnote{In this paper we call {\it physical} ({\it unphysical}) those states that (do not) respect level matching.} is encoded in the modular invariant partition function. Therefore, one can use properties of modular functions to gain a deeper understanding of string theory. Particularly useful to the problem at hand is an old result of Rankin and Selberg and its generalisation by Zagier that allows to connect the modular integral of generic automorphic functions, of at most moderate growth for $\tau_2 \to \infty$,
to the small $\tau_2$ behaviour of their constant term ({\it i.e.} $\tau_1$ independent) in the Fourier expansion. The difference of these two quantities has an oscillating behaviour whose frequencies are universally determined by the non-trivial zeroes of $\zeta (s)$. When applied to classically stable string theory vacua, this result implies that, not only the spectrum of physical excitations must enjoy {\it asymptotic supersymmetry} but actually, at very large mass, bosonic and fermionic states must follow a universal alternating pattern related to the zeros of the Riemann function, and independent of the string construction one is actually considering. Moreover, there is a strong correlation between the Riemann hypothesis and the rate of convergence of the total number of graded degrees of freedom to the vacuum energy, the convergence being fastest if the zeroes of $\zeta (s)$ are indeed on the critical line ${\rm Re}\, (s)=\frac{1}{2}$. We find this property remarkable.

Presently the strongest motivation for supersymmetry remains the stability it endows perturbatively to various string vacua. The results of 
\cite{Kutasov:1990sv} have changed the perspective since, in a class of string theories, exact supersymmetry is not needed to ensure the perturbative stability of the vacuum, but suffices to have an approximate {\it asymptotic supersymmetry}. The impact of this result was partially hampered by the feeling that, for large $N$, QCD has a Hagedorn-like spectrum, with purely bosonic excitations (also in the presence of matter at least if the number of colours is even) and without tachyons. As a result, the idea that QCD could be formulated as a string theory, perhaps in the presence of a RR background, and the absence of fermionic excitations in the large-$N$ spectrum, seems to be in contradiction with the results of \cite{Kutasov:1990sv}, and the thread relating supersymmetry and stability is cut. Fortunately, this is not automatically true, since stability does not universally imply {\it asymptotic supersymmetry}, as suggested by the existence of a large class of theories that, despite having no fermions in the spectrum, seem to be classically stable, even on a flat space-time without RR backgrounds.
These do not invalidate the Kutasov and Seiberg analysis \cite{Kutasov:1990sv} or the Rankin-Selberg-Zagier theorems \cite{RS,Zagier1,Zagier2}, since they correspond to (non-standard) orientifold projections of the type 0B theory, where the  closed oriented tachyon is odd under the orientifold group \cite{zerob}. In this case the IR divergence of the torus amplitude of type 0B theory is cancelled by the Klein bottle amplitude, and not by the presence of fermionic states that in these models are completely absent\footnote{at least in the closed-string sector. In the simplest instance of type $0'$ vacuum configuration an open string sector with both bosonic and fermionic excitations is demanded by RR tadpole cancellation, although examples can be devised in lower dimensions where unoriented closed strings do not require the introduction of D-branes.}. As a result {\it asymptotic supersymmetry} is no longer responsible for the finiteness of the vacuum energy and some new ultra-violet (UV) property of the string spectrum must be responsible for it. Unfortunately, both the presence of the physical tachyon in the parent oriented 0B string and the absence of modular symmetry in the Klein-bottle amplitude prevent from using properties of modular forms to draw meaningful conclusions. 

To recapitulate, we have found mathematical relations in number theory which allow us to elaborate on the IR/UV property relating stability to the graded high-mass density of states in the absence of physical closed oriented string tachyons. Although the physics suggests that such relation should persist also in the case of vacua with unphysical tachyons or configurations where the tachyon is removed by an orientifold projection, till now a convincing evidence evaded us. This makes the problem even more interesting and we hope to come back to it in the near future. Along similar lines it would also be interesting to apply recent results \cite{Cacciatori:2010js} on the equidistribution rate of modular forms of ${\rm Sp} (2g, {\mathbb Z})$ to genus-$g$ closed-string amplitudes and perhaps relate the finiteness of the higher-genus vacuum energy to properties of string interactions. Alternatively, one could use known results in string theory to shed light on mathematical aspects of automorphic forms as in \cite{cardella}.

The paper is organised as follows: in Section 2 we review some  properties of automorphic functions, and in particular the Rankin-Selberg method for computing modular integrals and its extension due to Zagier to the class of functions of moderate growth at the cusp. Section 3 contains some interesting applications to functions of interest in mathematics and string theory. In Section 4 we apply the mathematical properties outlined in Section 2 to describe the generic properties of string spectra and to quantities of physical interest. Sections 5 and 6 describe some concrete string examples. Finally, in Section 7 we study some specific heterotic vacua and give hints that the Rankin-Selberg-Zagier method might actually be extended also to the class of vacua with unphysical tachyons.

\section{The Rankin-Selberg-Zagier method for automorphic functions}
\label{rankinselberg}

Rankin and Selberg observed that there is a connection between the Mellin transform of the constant term in the Fourier expansion of an automorphic function and its inner product with an Eisenstein series \cite{RS}. 
As we shall see, this has consequences in the analysis of genus-one amplitudes in String Theory and, in a modified version, reproduces the results of Kutasov and Seiberg \cite{Kutasov:1990sv} on the counting of the graded number of degrees of freedom in oriented closed strings. 

\subsection{The case of functions of rapid decay at the cusp}

The Rankin-Selberg method \cite{RS} deals with automorphic functions $f (\tau_1 , \tau_2)$ of rapid decay at the cusp $\tau = i \infty$ of the $\varGamma = {\rm PSL} (2,\mathbb{Z})$ fundamental domain
\begin{equation*}
{\mathscr F} = \left\{ \tau \in \mathbb{H}\ |\ |\tau | > 1\,, -\tfrac{1}{2} <\tau_1 \le \tfrac{1}{2} \,\right\} \cup 
\left\{ \tau \in \mathbb{H}\ |\ |\tau | = 1\,, 0 \le \tau_1 \le\tfrac{1}{2} \,\right\}  \,,
\end{equation*}
where $\tau = \tau_1 + i \tau_2$ is the Teichm\"uller parameter of the world-sheet torus and $\mathbb{H}$ is the hyperbolic upper complex plane. For such functions, let us consider the integral
\begin{equation}
I (s) = \int_{\mathscr F} d\mu \, f(\tau_1 , \tau_2 ) \, E (\tau , s)\,, \label{RSintegral}
\end{equation}
with $d\mu= \tau_2^{-2}\, d\tau_1\, d\tau_2$ the $\varGamma$-invariant measure on ${\mathscr F}$, and $E ( \tau , s)$ the spectral Eisenstein series
\begin{equation*}
E ( \tau , s) = \sum_{\gamma \in \varGamma _\infty \backslash \varGamma} {\rm Im}\, (\gamma\cdot \tau )^s
= \tfrac{1}{2} \sum_{\substack{ c,d\in\mathbb{Z}\\  (c,d) =1}} \frac{\tau_2^s}{|c\tau + d|^{2s}}  
\qquad \quad
\left( \varGamma_\infty = \left\{ \left( \begin{matrix} 1 & n \\ 0 & 1 \end{matrix}\right) \,,\ n\in\mathbb{Z} \right\}\right)\,,
\end{equation*}
where $\gamma$ is a generic element in $\varGamma$ acting projectively on $\tau$.
Using the invariance of $f (\tau_1 , \tau_2)$ under the modular group, it is thus possible to unfold the integral (\ref{RSintegral})
\begin{equation}
\int_{\mathscr F} d\mu \, f(\tau_1 , \tau_2 ) \, E (\tau , s) = \int_0^\infty d\tau_2 \, \tau_2^{s-2}\, a_0 (\tau_2 ) = {\mathscr M} (\tau_2^{-1} a_0 (\tau_2), s) \,, \label{RStransform}
\end{equation}
and connect $I(s)$ to the Mellin transform $\mathscr{M}$ of the constant term
\begin{equation*}
a_0 (\tau_2 ) = \int_{-1/2}^{1/2}d\tau_1 \, f (\tau_1 , \tau_2 )
\end{equation*}
of the Fourier expansion $f =\sum_{n\in\mathbb{Z}} a_n (\tau_2 ) e^{2i\pi n \tau_1}$ of the automorphic function . One can deduce from eq. (\ref{RStransform}) that the Mellin transform, as a function of the complex variable $s$, inherits the same analytic properties of the Eisenstein series. In particular, since
\begin{equation*}
E^* (\tau , s) \equiv \zeta^* (2 s) \, E (\tau , s) = \pi^{-s}\, \varGamma (s) \, \zeta (2s) \, E (\tau , s)
\end{equation*}
is a holomorphic function in the $s$ complex plane, except at $s=0,1$ where it has simple poles, it follows that also $\zeta^* (2s)\, {\mathscr M} (\tau_2^{-1}\, a_0 (\tau_2) , s) $ can be meromorphically continued to all $s$, the only possible poles being at $s=0$ and at $s=1$. 

By the Reverse Mapping Theorem, if the singular part of the Mellin transform reads
\begin{equation*}
\varphi(s) \equiv {\mathscr M} (f,s) = \sum_{k,b} \frac{c_{k,b}}{(s-b)^{k+1}}+{\rm non\ singular}\,,
\end{equation*}
and if $f(x)$ is of rapid decay as $x\to\infty$, then 
\begin{equation}
f(x) = {\mathscr M}^{-1} (\varphi , x) \sim  \sum_{k,b} (-1)^k \, \frac{c_{k,b}}{k!} \, x^{-b}\, \log^k x   \label{RMT}
\end{equation}
for small $x$.

Applying these results to the Rankin-Selberg relation (\ref{RStransform}), one gets \cite{Zagier1, Zagier2}
\begin{equation}
a_0 (\tau_2 ) \sim \frac{3}{\pi} \int_{\mathscr F} d\mu \, f  + \sum_{\zeta^* (\rho) = 0} C_\rho \tau_2^{1-\rho/2}\qquad {\rm as} \quad \tau_2 \to 0\,, \label{RSconvrate}
\end{equation}
for smooth functions $f$ of rapid decay, where the sum is extended over the zeroes of $\zeta^* (s)$ or, alternatively, over the non-trivial zeroes of the Riemann $\zeta$-function, and $C_\rho$ are $f$-dependent constants. This equation actually states that the horocycle tangent to the cusp, ${\mathscr H}_{i\infty} (\tau_2) = {\mathbb R} +i\tau_2$ becomes equidistributed in the fundamental domain in the $\tau_2 \to 0$ limit, and therefore the average value of $f$ on long horocycles is equal to the average value of $f$ on the fundamental domain\footnote{A horocycle in the upper hyperbolic plane $\mathbb{H}$ is a circle tangent to rational points and to the cusp $\tau = i \infty$.
Every horocycle $\mathscr{H} (R)$ of radius $R$ has an image curve $\gamma_R$ under the action of the modular group $\varGamma$ that is fully contained inside the fundamental domain $\mathscr{F}$. In the infinite radius limit, $R\to \infty$, every horocycle 
converges to the same limiting curve, that is uniformly distributed in the fundamental domain $\mathscr{F}$.} 
\begin{equation*}
\lim_{\tau_2 \to 0} \langle \, f \, \rangle_{{\mathscr H}_{i\infty} (\tau_2 )} = \langle \, f \, \rangle_{\mathscr F}\,.
\end{equation*}
Generalisation of this relation to the case of equidistribution of horocycle flow in the modular domain of congruence subgroups of $\varGamma$ have been studied in \cite{Cardella:2008nz}, and are of interest for the study of orbifold compactification of closed strings.

\subsection{An extension to functions of moderate growth at the cusp}
\label{moderate}

These results themselves are not of direct applicability to String Theory, since typically genus-one integrands are at most of slow (power-law) decay at the cusp, if not exponentially divergent. 
In fact, typically string vacua always have massless excitations, that therefore contribute to the partition function with a term $d_0 \, \tau^\beta$, where $d_0$ counts the graded number of massless degrees of freedom and $\beta$ is a constant depending usually on the number of non compact dimensions $d$, $\beta = 1-d/2$. Therefore, the automorphic functions of interest in string theory generically are not of rapid decay, rather they are of moderate growth/decay at the cusp. Here we are not considering the case of vacua with physical and unphysical tachyons, corresponding to functions of rapid growth for which similar theorems are absent. We shall comment on this possibility in Section 7.

Fortunately, Zagier has provided a generalisation of the Rankin-Selberg method, applicable to the case of modular functions $f (\tau_1 , \tau_2)$ of moderate growth at the cusp \cite{Zagier2}\footnote{Due to the behaviour of the function at the cusp, the proof of the theorem in ref. \cite{Zagier2} involves a non-trivial unfolding procedure that also appears in \cite{Green:1999pv}, related to the evaluation of the low energy expansion of the genus one string amplitude.}. In particular, if as $\tau_2 \to \infty$ the automorphic function behaves as
\begin{equation*}
f(\tau_1 ,\tau_2 ) = \varphi (\tau_2 ) + O (\tau_2^{-N}) \qquad (\forall N>0)\,,
\end{equation*}
with
\begin{equation*}
\varphi (\tau_2 ) = \sum_{i=1}^\ell \frac{c_i}{n_i !} \, \tau_2^{\alpha_i} \, \log^{n_i} \tau_2\,,
\end{equation*}
he has shown that the Rankin-Selberg transform 
\begin{equation*}
R^* (f,s) \equiv \zeta^* (2s) \int_0^\infty d\tau_2\, \tau_2^{s-2} \left[ a_0 (\tau_2) - \varphi (\tau_2) \right]
\end{equation*}
can be meromorphically continued to all $s$, the only poles being at $s=0,\, 1,\, \alpha_i ,\, 1-\alpha_i$
\begin{equation}
R^* (f , s) = \sum_{i=1}^\ell c_i \left( \frac{\zeta^* (2s)}{(1-\alpha_i - s)^{n_i +1}} + \frac{\zeta^* (2s-1)}{(s-\alpha_i )^{n_i +1}} \right) + \frac{\varPhi (s) }{s(s-1)}\,, \label{zagier}
\end{equation}
with $\varPhi (s)$ an entire function. 

A direct use of the Reverse Mapping Theorem (\ref{RMT}) then implies that
\begin{equation}
\begin{split}
a_0 (\tau_2 ) &\sim \varphi (\tau_2) + C + \sum_{i=1}^\ell \frac{c_i}{n_i !} \left[ - \tau_2^{\alpha_i} + (-1)^{n_i}\, \varPsi (\alpha_i ) \, \tau_2^{1-\alpha_i }\right] \,  \log^{n_i} \tau_2  + \sum_{\zeta^* (\rho ) =0} C_\rho\, \tau_2^{1-\rho/2}
\\
&
= C + \sum_{i=1}^\ell \frac{ c_i}{n_i !}\, (-1)^{n_i}\, \varPsi (\alpha_i )
\,\tau_2^{1-\alpha_i}\, \log^{n_i}\tau_2 + \sum_{\zeta^* (\rho ) =0} C_\rho\, \tau_2^{1-\rho/2}
\end{split}
\label{a0}
\end{equation}
as $\tau_2 \to 0$. In this expression $C$ is a constant and $\varPsi (s) = \zeta^* (2s-1) / \zeta^* (s)$ is the automorphic factor that appears in the constant term of the Fourier development of the spectral Eisenstein series
\begin{equation*}
\int_{-1/2}^{1/2} d\tau_1 \, E ( \tau, s) = \tau_2^s + \varPsi (s)\, \tau_2^{1-s}\,, \qquad \left( s\not=0,\tfrac{1}{2},1 \right)\,.
\end{equation*}
Moreover, Zagier was able to show that the constant $C$ actually equals the average of $f$ over the fundamental domain
\begin{equation*}
C = \frac{3}{\pi} \int_{\mathscr F} d \mu \, f(\tau_1 , \tau_2 )\,,
\end{equation*}
whenever the integral converges. As we shall see in Section 4, eq. (\ref{zagier}) has important consequences in the mass distribution of degrees of freedom in type II superstring theories.

\section{Some applications of the Rankin-Selberg-Zagier method}
\label{examples}

Before we proceed to apply these results to string theory, it is instructive to consider a couple of interesting cases that can exemplify the power of the Rankin-Selberg-Zagier method. 

\subsection{Example one: functions of rapid decay}

Let us consider the automorphic function
\begin{equation*}
f (\tau_1 , \tau_2 ) = \tau_2^{12}\, |\varDelta (\tau ) |^2\,,
\end{equation*}
originally studied by Rankin and Selberg, where $\varDelta (\tau )$ is the discriminant function from the theory of elliptic functions
\begin{equation*}
\varDelta (\tau ) =   q\prod_{n=1}^\infty (1- q^n )^{24} = \sum_{n=1}^\infty \tau(n) \, q^n
\quad \qquad (q=e^{2i\pi \tau})\,,
\end{equation*}
with $\tau (n)$ the Ramanujan $\tau$-function. It follows that
\begin{equation*}
\displaystyle a_0 (\tau_2 ) = \displaystyle\tau_2^{12}\, \sum_{n=1}^{\infty} \tau^2 (n) e^{-4\pi n \tau_2}\,,\\[2mm]
\end{equation*}
while a numerical integration yields
\begin{equation*}
C=\frac{3}{\pi}\int_{\mathscr F} d\mu \, \tau_2^{12}\, |\varDelta (\tau ) |^2 \simeq 9.88698 \times 10^{-7}\,.
\end{equation*}
As a result, from eq. (\ref{RSconvrate}) we can conclude that, as $\tau_2 \to 0$,
\begin{equation*}
\tau_2^{12}\, \sum_{n=1}^{\infty} \tau^2 (n) e^{-4\pi n \tau_2} \sim C + \sum_{\zeta^* (\rho) =0} C_\rho \, \tau_2^{1-\rho/2} \,.
\end{equation*}
Assuming the Riemann hypothesis, {\it i.e.} $\rho = \frac{1}{2} + i \gamma$, the convergence rate is the fastest and the series oscillates around its limiting value with frequencies that depend on the location of the zeroes on the critical line ${\rm Re}\, (\tau )= \frac{1}{2}$
\begin{equation}
a_0 (\tau_2 ) \sim C + \tau_2^{3/4} \sum_{m=1}^\infty C_m \, \cos \left( \tfrac{1}{2} \, \gamma_m \log \tau_2 + \phi_m \right)\,,
\label{discriminant}
\end{equation}
where $C_m$ and $\phi_m$ are real constants. The graphs of $a_0 ( \tau_2 )$ and of $(a_0 (\tau_2 ) - C )/\tau_2^{3/4}$ are shown in fig. 1. The oscillations are evident and, doing a rough numerical Fourier analysis, Zagier found that the first two frequencies are $14.138/4\pi$ and $20.8/4\pi$ \cite{Zagier2}. These have to be compared to the first two non-trivial zeroes of the $\zeta$-function, $\gamma_1 \approx 14.135$ and $\gamma_2 \approx 21.0$. It is quite remarkable that eq. (\ref{discriminant}) can be used to evaluate the zeroes of the Riemann $\zeta$-function using the values of the Ramanujan $\tau$-function.

\medskip

\begin{figure}[h] 
	\begin{center}
		\includegraphics[scale=.75]{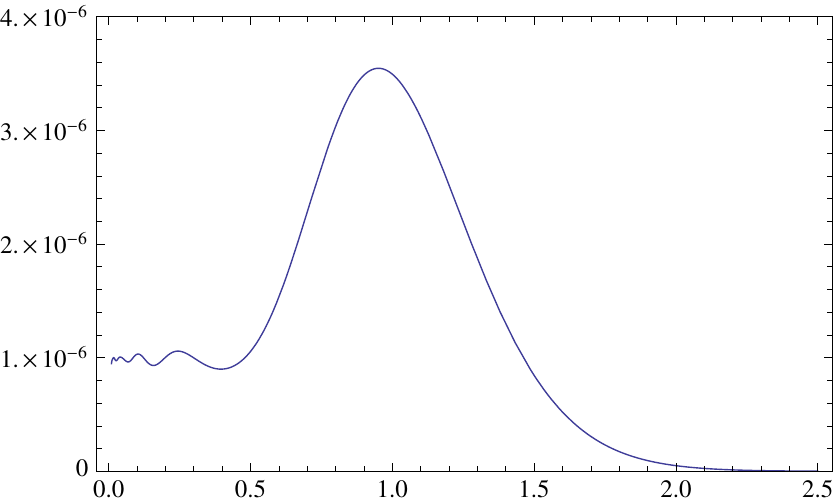} \hskip 20pt
		\includegraphics[scale=.75]{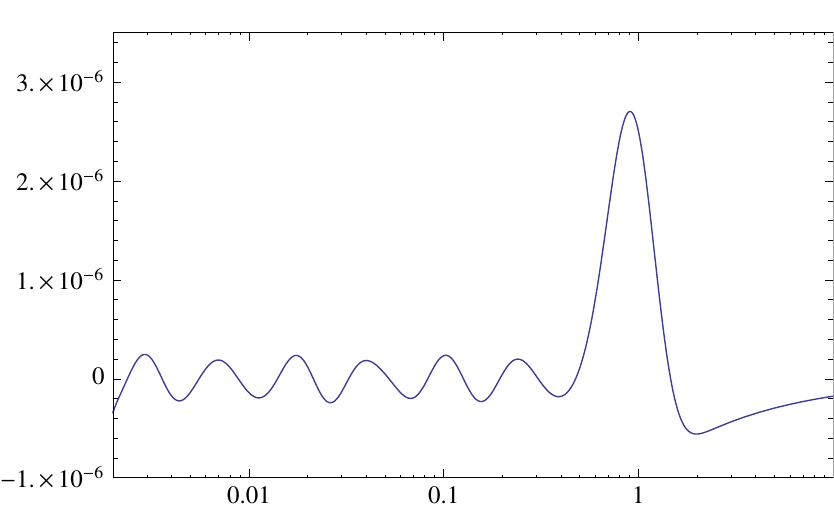} 
		\caption{\small In the left figure we plot the constant term of the discriminant function $\tau_2^{12} |\varDelta (\tau) |^2$, as $\tau_2 \to 0$. The right figure represents the error term $(a_0 (\tau_2 ) - C)\, \tau_2^{-3/4}$and clearly shows the oscillations due to the non-trivial zeroes of the Riemann $\zeta$-function.} 
	\end{center}
\end{figure}

\bigskip

\subsection{Example two: functions of moderate growth}

The next example we want to study is the case of
\begin{equation*}
f(\tau_1 , \tau_2 ) = R\, \varTheta (\tau , R ) = R\, \sum_{m,n \in \mathbb{Z}} e^{- \frac{\pi R^2}{\tau_2} |m \tau + n |^2}\,,
\end{equation*}
where $R$ is a (real) parameter and $\varTheta  (\tau , R )$ is the Poincar\'e series of the theta series $\vartheta (R^2 /\tau_2 )$, with
\begin{equation*}
\vartheta (t) = 2 \, \sum_{r=1}^{\infty} e^{-\pi r^2 t}\,.
\end{equation*}

From a string theory perspective, the modular invariant function $\varTheta (\tau , R)$ encodes the contribution of the zero modes of a free boson compactified on a circle of radius $R$ to the one-loop partition function. After a Poisson resummation it can be written as
\begin{equation*}
R\, \varTheta (\tau , R) = \sqrt{\tau_2}\, \sum_{m,n\in\mathbb{Z}} q^{\frac{1}{4} \left( \frac{m}{R} + nR\right)^2}\, \bar q ^{\frac{1}{4} \left( \frac{m}{R} - nR\right)^2} \,,
\end{equation*}
from which it follows that $ f (\tau_1 ,\tau_2 )$ is a function of moderate growth at the cusp
\begin{equation*}
R \, \varTheta (\tau , R ) \sim \sqrt{\tau_2} + o ( \tau_2^{-N})\qquad (\forall N>0)\qquad {\rm as}\quad \tau_2 \to \infty\,.
\end{equation*}
One of the features of $f (\tau_1 , \tau_2 )$ is that its average on the fundamental domain can be computed analytically 
using the properties of the Poincar\'e series
\begin{equation}
\int_{\mathscr F} d\mu\, R\, \varTheta (\tau , R ) = R \int_{\mathscr F} d\mu + R \, {\sum_r } ' \int_0^\infty \frac{d\tau_2  }{\tau_2^2}\, e^{-\pi r^2 R^2/\tau_2} = \frac{\pi}{3} \left( R +\frac{1}{R} \right)\,,
\label{Thetaint}
\end{equation}
where, as usual, a primed sum does not include the term $r=0$.
Following \cite{Zagier2}, from this expression one can read that $\lim_{\tau_2 \to 0} a_0 (\tau_2 ) = R + R^{-1}$, and indeed a direct calculation yields
\begin{equation}
\begin{split}
a_0 (\tau_2 , R) &= \int_{-1/2}^{1/2} d\tau_1\, R\, \varTheta (\tau , R) = \sqrt{\tau_2 } \, \sum_{m,n\in \mathbb{Z}} e^{- \pi \tau_2 \left(\frac{m^2}{R^2} + n^2 R^2 \right)} \int_{-1/2}^{1/2} d\tau_1\, e^{2 i \pi \tau_1 \, m n}
\\
&= R + R^{-1}- \sqrt{\tau_2 } + {\sum_{m\in \mathbb{Z}}}^\prime \left( R\, e^{-\pi R^2 m^2/\tau_2} + R^{-1} \, e^{-\pi  m^2/R^2\tau_2}  \right)
\,.
\end{split}
\label{Thetaapprox}
\end{equation}
Figure 2 shows the exact expression (\ref{Thetaint}) versus the constant term $a_0 (\tau_2 , R)$ as a function of $R$ for various values of $\tau_2$ small.

\medskip

\begin{figure}[h] 
	\begin{center}
		\includegraphics[scale=1]{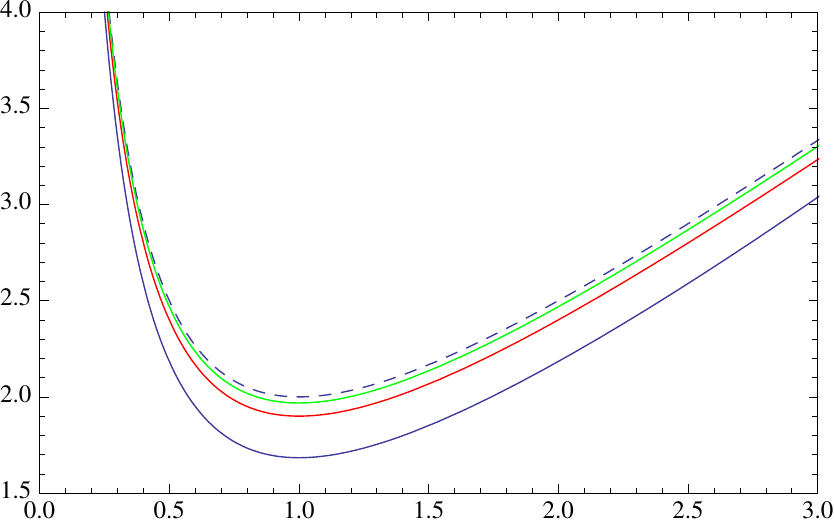}
		\caption{\small A comparison between the exact value of average of $\varTheta (\tau , R)$ over the fundamental domain and the constant term $a_0 (\tau_2 , R)$ as a function of $R$ for various values of $\tau_2$. The dashed line corresponds to $3\, \pi^{-1}\,\langle \varTheta \rangle_{\mathscr F}$, the solid blue line to $a_0 (0.1,R)$, the red line to $a_0 (0.01,R)$ and, finally, the green line to $a_0 (0.001,R)$. The approximation becomes more accurate as $\tau_2$ decreases, and $a_0$ approaches $3\, \pi^{-1}\,\langle \varTheta \rangle_{\mathscr F}$ monotonically.
		} 
	\end{center}
\end{figure}

From eq. (\ref{Thetaapprox}) it is evident that $a_0 (\tau_2 , R)$ approaches monotonically the exact average $3\, \pi^{-1}\, \langle \varTheta \rangle_{\mathscr F}$ since it is the sum of positive terms. 
The expected oscillations induced by the non-trivial zeroes of the Riemann $\zeta$-function seem not to play any role in this example. In fact, taking into account that in our case $\varphi (\tau_2 ) = \sqrt{\tau_2}$, and thus $\ell =1$, $\alpha_i =\frac{1}{2}$, $c_i =1$, $n_i =0$, from eq. (\ref{a0}) one gets
\begin{equation*}
\begin{split}
a_0 (\tau_2 ) &\sim \sqrt{\tau_2 } + C - \left(1-\varPsi (\tfrac{1}{2} ) \right) \sqrt{\tau_2} + \sum_{\zeta^* (\rho ) =0} C_\rho \, \tau_2^{1-\rho/2}
\\
&\sim R + R^{-1} - \sqrt{\tau_2} + \sum_{\zeta^* (\rho ) =0} C_\rho \, \tau_2^{1-\rho/2}\,,
\end{split}
\end{equation*}
and $\sqrt{\tau_2}$ overrides $\tau_2^{1-\rho/2}$, for small values of $\tau_2$, since it is known that for the non-trivial zeroes of the Riemann $\zeta$-function  $\frac{1}{2} \le \sup {\rm Re}\, (\rho  )<1$.

Actually, it is possible to prove that for this example the coefficients $C_\rho $ are vanishing identically. Let us consider, in fact, the Rankin-Selberg integral
\begin{equation*}
I (R,s) = \int_{\mathscr{F}} d\mu\, \varTheta (\tau ,R )\, E (s,\tau )\,.
\end{equation*}
As already anticipated, $I(R,s )$ inherits the analytic properties of the Eisenstein series, and thus is expected to have a simple pole at $s=\frac{1}{2} \rho $ with residue
\begin{equation*}
C_\rho (R) = {\rm Res}_{s=\rho/2} \, I (R,s)\,,
\end{equation*}
or, alternatively,
\begin{equation*}
C_\rho (R) = \int_{\mathscr{F}}d\mu\, \varTheta (\tau , R ) \, E^* (\tfrac{1}{2}\,\rho,\tau )\,.
\end{equation*}
$C_\rho$ can actually be computed explicitly by noting that the spectral Eisenstein series is the Mellin transform of the Theta series, or equivalently the Theta series is the inverse Mellin transform of $E (s,\tau )$
\begin{equation*}
\varTheta (\tau , R ) = \frac{1}{2\pi i }\int_{c-i\infty}^{c+i\infty} ds\, R^{-s}\, E^* (s,\tau )\,,
\end{equation*}
and the $E (s,\tau )$ are eigenfunctions of the Laplace operator on the hyperbolic upper-complex-plane and thus obey orthonormality conditions
\begin{equation*}
\int_{\mathscr{F}}d\mu\, E (s , \tau )\, E (s', \tau ) = \delta (s-s')\,.
\end{equation*}
Using these results, one gets
\begin{equation*}
C_\rho (R) = \frac{1}{2\pi i} \, R^{-\rho/2}\, [\zeta^* (\rho ) ]^2 \equiv 0\,,
\end{equation*}
since $\rho$ is a zero of $\zeta^* (s)$.

\section{Constraints on the density of states in non-supersym\-metric Type II superstrings}

After having introduced the main mathematical tools in Section \ref{rankinselberg}, we can now move to discuss their implications for the density of states in non-supersymmetric non-tachyonic superstrings. In supersymmetric theories the spectrum of excitations is Fermi-Bose degenerate at each mass-level and therefore the partition function is vanishing identically, while tachyonic vacua are classically unstable and yield a divergent one-loop amplitude.

The quantity we want to study is
\begin{equation*}
\varOmega =\int_{\mathscr{F}} \frac{d^2 \tau}{\tau_2^2} \, \mathscr{Z}_{\rm string}\,,
\end{equation*}
where 
\begin{equation}
\mathscr{Z}_{\rm string} = {\rm Str}_{\mathscr{H} \times\bar {\mathscr{H}}}\, q^{L_0 - E_0}\, \bar q^{\bar L_0 - \bar E_0} \simeq
\tau_2^{1-d/2 }
\sum_{\varDelta , \bar\varDelta} \, c(\varDelta , \bar\varDelta)\, q^{\varDelta - E_0} \, \bar q^{\bar\varDelta - \bar E_0}\,,
\label{part}
\end{equation}
where $d$ denotes the number of non-compact directions, $\varDelta, \, \bar\varDelta$ are the conformal dimensions of the (anti)holomorphic string states, while $E_0$ and $\bar E_0$ denote the energy of the left-handed and right-handed vacua where, for instance, in flat space-time a single periodic boson contributes $-\frac{1}{24}$, while a single fermion contributes $-\frac{1}{48}$ if it is anti-periodic and $\frac{1}{24}$ if it is periodic. Finally, the integers $c (\varDelta , \bar\varDelta)$ count the (signed) number of states with dimension $ (\varDelta , \bar\varDelta)$, and are positive for bosons and negative for fermions. 

The torus amplitude $\varOmega$ probes the stability of the string vacuum described by $\mathscr{Z}_{\rm string}$, and computes the radiative correction to the cosmological constant. In some interesting compactifications, the Virasoro operators $L_0$ and $\bar L_0$ depend on the geometric moduli describing the size and shape of the internal manifold, and therefore the Casimir energy actually generates a Coleman-Weinberg-like potential that can lead to perturbative moduli stabilisation \cite{Angelantonj:2006ut,Angelantonj:2008fz}. 

If tachyons are present in a given string vacuum, or if they appear in some regions of its moduli space, the vacuum energy is exponentially divergent and the string configuration is unstable. In addition, the emergence, at finite temperature, of tachyons and the divergence of $\varOmega$ signal the onset of first-order phase transitions \cite{Atick:1988si,ridge}, and therefore the perturbative description of the vacuum is not valid any longer. 
That is why in the following we shall restrict our attention to the case of non-tachyonic string vacua, eventually limiting our analysis to those regions in moduli space where the mass-spectrum of the theory is semi-positive definite. 

Actually, the theory may have so-called unphysical tachyons, with $\varDelta + \bar\varDelta -E_0 -\bar E_0 <0$ but $\varDelta -E_0 \not = \bar\varDelta -\bar E_0$. In such a situation, $\varOmega$ depends on the way integration on $\mathscr{F}$ is performed. One is instructed to integrate first over $\tau_1$ and then over $\tau_2$, for large $\tau_2$ \cite{GSW}. This amounts to impose first level-matching, and therefore, for large $\tau_2$, only physical states --- those with $\varDelta -E_0 = \bar\varDelta -\bar E_0$ --- contribute to the vacuum energy, that is now finite. Although in this case $|\varOmega | <\infty $ and thus the string vacuum is classically stable, the presence of unphysical tachyons invalidates the analysis of Rankin-Selberg-Zagier, and therefore in the following we shall consider only configurations where both physical and unphysical tachyons are absent\footnote{Notice that it is not true in general that the vacuum energy receives contributions only from the physical states, since functions with vanishing constant term in their Fourier expansion can be integrated over the fundamental domain yielding a finite non-zero result. For instance the $j$ cusp form clearly does not satisfy level matching, but $\int_\mathscr{F} d\mu \, j(\tau ) = - 8 \pi $, the contribution coming entirely from the region $\mathscr{L} =\{ \tau\in\mathbb{H}\ | |\tau|\ge 1\,, -\tfrac{1}{2} <\tau_1 \le \tfrac{1}{2}\,, 0<\tau_2 \le 1 \}$.}. We shall come back to these classes of vacua in section 7.

Under these conditions, the string partition function $\mathscr{Z}_{\rm string}$ is an automorphic function of moderate decay at the cusp\footnote{Actually, if the number of non-compact dimensions is $d\le 1$ the partition function if of moderate growth. Such cases typically emerge in the analysis of early cosmological evolutions \cite{cosmology}.}, and therefore one can directly apply the results of section \ref{moderate} to the vacuum energy $\varOmega$. Since the only term of moderate decay at $i\infty$ is
\begin{equation*}
\varphi (\tau_2 ) = c(E_0 , \bar E_0 ) \, \tau_2^{1-d/2} \equiv d(0)\, \tau_2^{1-d/2}\,,
\end{equation*}
one gets the relation
\begin{equation*}
a_0 (\tau_2 ) \sim \frac{3}{\pi} \, \varOmega + d (0) \,  \varPsi (1-\tfrac{1}{2}\, d) \, \tau_2^{d/2} + \sum_{\zeta^* (\rho) =0} C_\rho \, \tau_2^{1-\rho/2}
\,,
\end{equation*}
valid for very small $\tau_2$.

It is actually useful to identify the proper time $\tau_2$ with an inverse UV cut-off, $\tau_2 = \varLambda^{-2}$, and define the function
\begin{equation*}
\begin{split}
g (\varLambda ) &= \varLambda^{2-d} \, a_0 (\varLambda^{-2} ) 
\\
&=\sum_{\varDelta , \bar\varDelta} c(\varDelta , \bar\varDelta ) \,e^{-2 \pi (\varDelta + \bar\varDelta - E_0 - \bar E_0)/\varLambda^2}\, \delta_{\varDelta , \bar\varDelta}
\\
&\equiv \sum_{\{ m^2\}} d ( m^2 ) \, e^{-4 \pi m^2  /\varLambda^2} \,,
\end{split}
\end{equation*}
where the sum in the third line is extended to all string states with mass $m^2  = \frac{1}{2} (\varDelta +\bar \varDelta - E_0 -\bar E_0)$ and  $d(m^2 ) \equiv d_{\rm B} (m^2 ) - d_{\rm F} (m^2)= c (\varDelta , \bar\varDelta )\, \delta_{\varDelta - E_0 , \bar \varDelta -\bar E_0}$ counts the graded number of physical degrees of freedom at a given mass. Eq. (\ref{a0}) results in 
\begin{equation}
g (\varLambda ) \sim \frac{3}{\pi}\,\varLambda^{2-d}\, \varOmega + \varLambda^{-d} \sum_{\zeta^* (\rho) =0} C_\rho \,\varLambda^{\rho} + d(0)\, \varPsi (1-\tfrac{1}{2} d)\, \varLambda ^{2-2d} \,,
\label{main}
\end{equation}
now valid for large values of the cut-off $\varLambda$. Since $\frac{1}{2} \le \sup {\rm Re}\, (\rho ) <1$, the first term is dominating for large values of $\varLambda$, while the last can be consistently neglected, at least in the case of $d>1$ non compact dimensions. Notice, however, that the minimal error, obtained in the case the Riemann hypothesis is true and ${\rm Re}\, (\rho) = \frac{1}{2}$, is proportional to $\varLambda^{1/2 -d}$ and therefore the convergence rate of $g(\varLambda )$ to the exact value of the vacuum energy is extremely slow. 

This equation has a twofold interpretation. On the one hand, it constrains the graded density of states in a classically stable string vacuum while, on the other hand, it is a way to well approximate the vacuum energy by simply summing a finite number of string states\footnote{though large! See the discussion in Section \ref{temperature} on the slow convergence rate of the series.}.

\medskip

\begin{figure}[h] 
	\begin{center}
		\includegraphics[scale=.5]{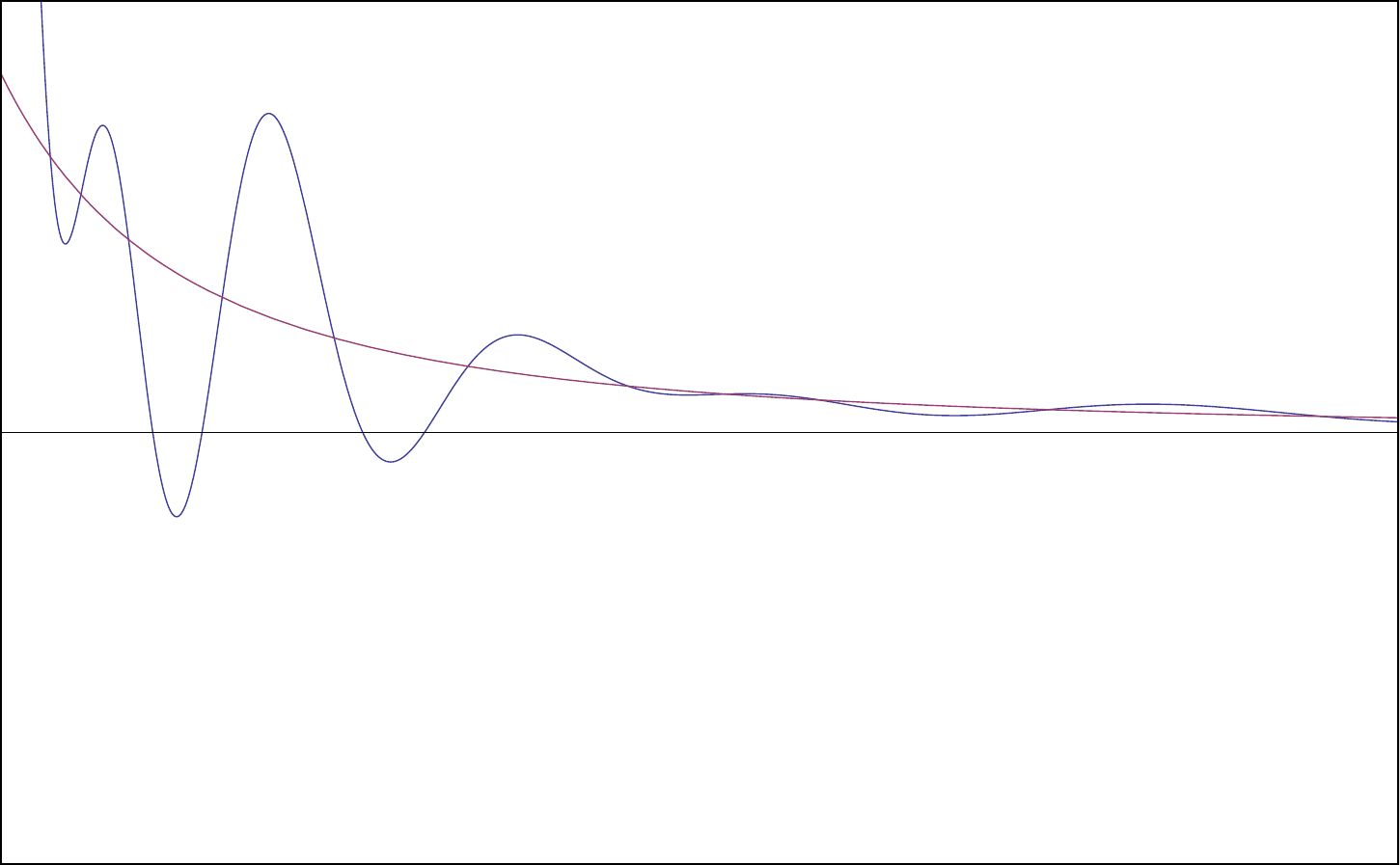}
		\caption{\small A generic behaviour of eq. (\ref{main}). The red line corresponds to the main contribution linked to the vacuum energy $\varOmega$, while the wavy blue lines includes the (oscillating) error terms.
		} 
	\end{center}
\end{figure}

In fact, from eq. (\ref{main}) one recovers that $g (\varLambda )$ vanishes
\begin{equation*}
\lim_{\varLambda \to\infty} g (\varLambda ) = \lim_{\varLambda \to\infty} \left[
\frac{3}{\pi}\, \varOmega \, \varLambda^{2-d} + \sum_{\zeta^* (\rho) =0} C_\rho \,\varLambda^{\rho-d} \right] =0\,,
\end{equation*}
for more than two non compact dimensions. This is the result of  \cite{Kutasov:1990sv} implying {\it asymptotic supersymmetry} in a classically stable vacuum. 
However, one has derived in addition a more stringent constraint on the allowed deviations from supersymmetry in a tachyon-free oriented superstring theory. For example, assuming the Riemann hypothesis, one can write
\begin{equation*}
\sum_{\zeta^* (\rho )=0} C_\rho \varLambda^{\rho -2} = \varLambda^{-3/2}\, \sum_m C_m \,\cos \left( \gamma_m \, \log \varLambda + \phi_m \right) \,,
\end{equation*}
where we have put $\rho = \frac{1}{2} +i\gamma$ and the sum is extended over the zeros on the upper complex plane, $\gamma >0$.
This implies that $g (\varLambda )$ oscillates about its limiting value $\frac{3}{\pi} \varOmega \varLambda^{2-d}$, as is schematically shown in fig. 3. The graded density of states at a given mass has an alternating sign behaviour as a function of the mass. The frequency of the oscillations are dictated by the zeroes of the Riemann $\zeta$-function; we find this remarkable\footnote{It is amusing to notice that the Riemann hypothesis appears to be correlated to other physical problems in quantum physics \cite{deRafael:2010ac}.}. Following a different analysis, Dienes \cite{Dienes:1994es} already noted that the graded density of states has a subtle oscillatory behaviour, and he called this phenomenon {\it misaligned supersymmetry}.

By inverting eq. (\ref{main})
\begin{equation*}
\varOmega \sim \frac{\pi}{3}\, \varLambda^{d-2} \, \sum_{\{m^2\}} d (m^2 )\, e^{-4\pi m^2 /\varLambda^2} + O( \varLambda^{\rho_{\rm sup} -2}) \qquad \left( \rho_{\rm sup} = \sup_{\zeta^* (\rho) =0} \, ({\rm Re}\,  (\rho)) \right)\,,
\end{equation*}
one can also use this expression in a more practical way to approximate the vacuum energy of a given string vacuum by just adding a finite number of degrees of freedom, since very massive states essentially do not contribute to $\varOmega$. In principle, this is a great simplification since $\varOmega$ involves a complicated integral over the fundamental domain, and typically unfolding techniques do not always allow to disentangle the $\tau_1$ and $\tau_2$ integrals since a $q$-expansion  of $\mathscr{Z}_{\rm string}$ might not be uniformly convergent for any value of $\tau_2$ \cite{Angelantonj:2006ut}. However, as we shall discuss in Section \ref{temperature} the convergence of the result is a delicate issue, since typically one needs an almost exact cancellation among extremely large numbers.

The Rankin-Selberg-Zagier analysis is not only relevant for the computation of the vacuum energy and/or for the counting of the graded number of degrees of freedom. It can be straightforwardly extended to constrain the supertrace of the operator $M^{2n}$, $M^2$ being the squared-mass-matrix. ${\rm Str}\, M^{2n}$ is of great interest since it controls the pattern of UV divergences of loop amplitudes in field theory. Recalling the definition
\begin{equation*}
g (\varLambda ) = \sum_{\{ m^2\}} d (m^2) e^{-4\pi m^2 /\varLambda^2} = {\rm Str} \, \left( e^{-4\pi M^2 /\varLambda^2 }\right) \,,
\end{equation*}
one can therefore use $g (\varLambda )$ as a generating function for mass superstrace
\begin{equation*}
{\rm Str} \left( M^{2k} \, e^{-4 \pi M^2 /\varLambda^2}\right) = \frac{(-1)^k}{(4\pi)^k}\, \frac{d^k}{d (\varLambda^{-2})^{k}} 
{\rm Str} \, \left( e^{-4\pi M^2 /\varLambda^2 }\right) \,.
\end{equation*}
Our main equation (\ref{main}) can be used to further constrain the way supersymmetry can be broken in a classically stable string vacuum
\begin{equation}
\begin{split}
{\rm Str}\, \left( M^{2k}\, e^{-4 \pi M^2 /\varLambda^2}\right) =&\, \frac{(-1)^k}{(4\pi )^k} \left[ \left( \tfrac{1}{2} d -2 \right)_k
\,\varOmega\, \varLambda^{2k+2-d} \right.
\\
&  + \sum_{\zeta^* (\rho)=0} \left( \tfrac{1}{2} (d-\rho) -1 \right)_k \, C_\rho \, \varLambda^{2k+\rho-d}
\\
&\left.  + d (0)\, ( d-2 )_k \, \varPsi (1-\tfrac{1}{2}\, d)  \, \varLambda^{2k+2-2d }\right]
\end{split}
\label{superstrace}
\end{equation}
where
\begin{equation*}
(x)_k = \frac{\varGamma (x+1 ) }{\varGamma ( x+1-k )}
\end{equation*}
 is the Pochhammer symbol, and indicates the falling sequential product. The first term dominates for large cut-off and, in $d$ non compact space-time dimensions,
\begin{equation*}
{\rm Str} \, M^{2k} = \lim_{\varLambda \to \infty} {\rm Str}\, \left( M^{2k}\, e^{-4 \pi M^2 /\varLambda^2}\right) =0 \quad {\rm for} \qquad k<\tfrac{1}{2}d -1\,.
\end{equation*}
Also in this case, eq. (\ref{superstrace}) spells-out how the infinite tower of string states conspire to yield a vanishing ${\rm Str} \, M^{2k}$.

\section{Application to finite temperature superstrings}
\label{temperature}

Type II superstrings at finite temperature represent the simplest instance where the Fermi-Bose degeneracy at $T=0$ is broken in a controlled way, at least for small enough $T$. As in Field Theory, bosons are periodic in the compact Euclidean time while fermions are anti-periodic and therefore have shifted Kaluza-Klein masses, thus breaking the Fermi-Bose degeneracy of the spectrum. Moreover, closed strings can wind arbitrary number of times around the time-circle, and a non-vanishing $T$ affects the GSO projection of the $T=0$ theory, so that sectors with odd windings have opposite GSO phases. This implies that, for $T\not = 0$, string perturbation theory is valid only for low temperatures, since above the critical value $T_{\rm H} = 2/\sqrt{\alpha '}$ tachyons appear in the spectrum, thus signalling the onset of a first-order Hagedorn transition \cite{Atick:1988si}. 

The finite-temperature effects on the whole string spectrum are encoded as usual in the partition function
\begin{equation*}
\begin{split}
\mathscr{Z}_T &=\frac{ 1}{ \tau_2^{7/2}\, (\eta \,\bar\eta )^8}\, \sum_{m,n\in\mathbb{Z}} \left[ \left( V_8 \,\bar V_8 + S_8 \bar S_8 \right) \, \varGamma_{m,2n} (T) - \left( V_8 \, \bar S_8 +S_8 \,\bar V_8 \right)\, \varGamma_{m+\frac{1}{2},2n} (T) \right.
\\
& \left. + \left( O_8\, \bar O_8 + C_8 \,\bar C_8 \right) \,\varGamma_{m,2n+1} (T) - \left( O_8\, \bar C_8 + C_8 \,\bar O_8 \right)\, \varGamma_{m+\frac{1}{2},2n+1} (T) \right]\,, 
\end{split}
\end{equation*}
where $O_8$, $V_8$, $S_8$ and $C_8$ denote the level-one SO(8) characters \cite{Angelantonj:2002ct}, while
\begin{equation*}
\varGamma_{m+a,n+b} (T) = q^{\frac{\alpha ' }{4} ((m+a)T + (n+b)/\alpha' T )^2} \, \bar q^{\frac{\alpha ' }{4} ((m+a)T - (n+b)/\alpha' T )^2}
\end{equation*}
encodes the contribution of the (shifted) momenta and windings associated to the Euclidean time compactified on a circle of radius $T^{-1}$. 

At low temperature, the vacuum energy density
\begin{equation*}
\varOmega = \int_\mathscr{F} \frac{d^2 \tau}{\tau_2^2} \, \mathscr{Z}_T (\tau_1 , \tau_2 )
\end{equation*}
 can be computed using standard unfolding techniques to yield
\begin{equation}
\varOmega =  \frac{T^9 \, (\alpha ')^{9/2}\, d_0^2 \,\varGamma (5)\, \zeta (10,\tfrac{1}{2})}{2^9 \,\pi^5} + 4\, (\alpha ')^2 \,T^4 \sum_{N=1}^\infty \sum_{p=0}^\infty \frac{N^{5/2} \, d_N^2}{(p+\tfrac{1}{2} )^5}\, K_5 \left( \frac{8 \pi (p+\tfrac{1}{2}) \sqrt{N}}{T\sqrt{\alpha '}}\right)\,.
\label{vacbessel}
\end{equation}
Here $d_N$ are the coefficients in the expansion
\begin{equation*}
\frac{\vartheta_2^4 (0|\tau) }{\eta^{12} (\tau)} = \sum_{N=0}^\infty d_N \, q^N\,,
\end{equation*}
$\zeta (s,a)$ is the Hurwitz $\zeta$-function, and 
\begin{equation*}
K_n (2 \sqrt{x}) = \tfrac{1}{2}\, x^{- n/2}\, \int_0^\infty dt\, t^{n-1}\, e^{-  t - x/t}
\end{equation*}
is the modified Bessel function of the second kind.

The Bessel functions are exponentially suppressed for large values of their argument, and therefore for small temperatures they give a negligible contribution to $\varOmega$, that is thus strongly dominated by the massless states. Notice that in our case $d_0 =2^7$ (the total number of massless bosons, 64 in the NS-NS sector and 64 in the R-R sector) and therefore
\begin{equation*}
\varOmega (T) \simeq 10^4\,(T\, \sqrt{\alpha '})^9 \,,
\end{equation*}
As a result, $\varOmega$ is rather small, since the expression (\ref{vacbessel}) is only valid for $T<1/\sqrt{\alpha'}$. For instance, $\varOmega \simeq 10^{-5}$ for $T=1/(10 \,\sqrt{\alpha '})$.

Alternatively, according to eq. (\ref{main}), $\varOmega$ can be accurately evaluated by adding a finite number of string states, depending on the value of the cuf-off $\varLambda$,
\begin{equation*}
\varOmega \simeq \frac{\pi}{3} \,\varLambda^7 \, g (\varLambda ) \simeq \frac{\pi}{3} \, \varLambda^7 \, \sum_{\{m^2\}} d (m^2) \, e^{-4\pi m^2 /\varLambda^2} = \frac{\pi}{3}\, \varLambda^7 \sum_{\{m^2\}} \left[d_{\rm B} (m^2) - d_{\rm F} (m^2) \right] \, e^{-4\pi m^2 /\varLambda^2}\,.
\end{equation*}
The temperature dependence is now hidden in the definition of $d(m^2 )$ that now does not correspond to a ``real'' density of states 
\cite{ridge}.
In general, this expression represents a great simplification with respect to the evaluation of one-loop integrals, since it is much easier to Taylor expand a function rather than to numerically compute intertwined bi-dimensional integrals\footnote{Although in this simple case, the integral of the fundamental domain can be easily unfolded and it is possible to express the result in terms of special functions, this is not always the case when dealing with more complicated models.}. This was the case for the first example discussed in Section \ref{examples}. From the multiplicative formula $\tau (m) \tau (n) = \sum_{d|(m,n)} d^{11}\, \tau (mn/d^2)$, and the inequality
\begin{equation*}
|\tau (p)| < 2\, p^{11/2} \quad {\rm for}\ p\ {\rm prime},
\end{equation*}
one can deduce that the Ramanhujan $\tau$-function is of moderate growth. As a result, the convergence rate of the series $\sum_{n=1}^\infty \tau ^2 (n) \, e^{-4\pi n \tau_2}$ is extremely fast, and it is enough to sum only few terms to correctly estimate the modular integral of $\tau_2^{12} |\varDelta |^2$.

When applied to string theory, however, the convergence rate of the series becomes extremely slow, and one needs to sum a very large number of terms. This is due essentially to the term $|\eta (\tau ) |^{-24}$ ubiquitously present in the string partition function, that is responsible for the celebrated exponential growth of degrees of freedom in string theory. Indeed, if we Taylor expand the combinations
\begin{equation*}
\frac{\vartheta^4_\alpha (0|\tau ) }{\eta^{12} (\tau )} = \sum_{n=0}^\infty d_{\alpha\,,\, n} \, q^{n+h_\alpha }\,,
\end{equation*}
the Hardy-Ramanhujan formula yields an almost universal dominant behaviour
\begin{equation*}
d_{\alpha \,, \, n} \sim A_\alpha \, n^{-B_\alpha}\, e^{\pi \sqrt{8\, n}}\,,
\end{equation*}
with $A_\alpha$ and $B_\alpha$ constants. As a result, a sum like
\begin{equation*}
\frac{1}{2}\sum_{N=0}^\infty d^2_{2,N} \, e^{-4 \pi N /\varLambda^2}
\end{equation*}
converges only after one sums all the degrees of freedom up to the $N_{\rm max} \sim \sqrt{2} \,\varLambda^4$ level. Moreover, one has to carefully truncate the sum since two huge numbers should cancel with extremely high accuracy, since their difference must yield the very small number $\varOmega$. The situation is much more complicated when one is dealing with string vacua with some directions compactified, for instance, on irrational tori, and sums over multi-dimensional discrete momenta and windings have to be taken into account. Extreme care is therefore needed in summing the degrees of freedom and often powerful computational tools are needed.

\section{Scherk-Schwarz dimensional reduction on a rational ${\rm SO} (8)$ lattice}

To simplify the calculability of the model, we now consider the Scherk-Schwarz dimensional reduction of type II superstrings on a rational ${\rm SO} (8)$ lattice. Scherk-Schwarz deformations and finite temperature effects are closely related, since in both cases one is deforming the periodicity conditions of fields along compact directions, space-like in the Scherk-Schwarz case and (Euclidean) time-like in the finite temperature case. 

In the model we shall study, the Scherk-Schwarz deformation can be easily implemented by computing an asymmetric $\mathbb{Z}_2$ orbifold of the supersymmetric theory, where the $\mathbb{Z}_2$ generator is $g= (-1)^F \, I_{\rm L}$, where $F$ is the space-time fermion number and $I_{\rm L}$ the asymmetric inversion of the compact coordinates, $X_{\rm L} \to - X_{\rm L}$ and $X_{\rm R} \to + X_{\rm R}$. 
In terms of the level-one ${\rm SO} (8)$ characters \cite{Angelantonj:2002ct} the partition function reads
\begin{equation*}
\begin{split}
\mathscr{Z} =& \tau_2^{-2}\, (\eta \,\bar \eta )^{-4}\, \Bigl[ \left(  | V_8 |^2 + |S_8 |^2 \right) \, \left( |O_8 |^2 + |S_8|^2 \right) + \left( |O_8 |^2 + |C_8 |^2 \right)\,
\left( V_8 \, \bar C_8 + C_8 \, \bar V_8 \right)
\\
&-  \left(V_8 \, \bar S_8 + S_8 \, \bar V_8 \right) \, \left( |V_8 |^2 + |C_8|^2 \right) - \left( O_8 \, \bar C_8 + C_8 \, \bar O _8 \right) \, \left( O_8 \, \bar S_8 + S_8 \, \bar O _8 \right) \Bigr]
\end{split}
\end{equation*}
and we can deduce that the spectrum of light excitations, given in terms of representations of the ten-dimensional little group ${\rm SO} (8)$, comprises a graviton, a dilaton and a two-form from the ${\rm NS}_+$--${\rm NS}_+$ sector, a scalar, a two-form and a self-dual four-form from the ${\rm R}_+$--${\rm R}_+$ sector, 128 scalars from the ${\rm NS}_-$--${\rm NS}_-$ sector, and 16 right-handed Majorana-Weyl fermions from the ${\rm NS}_-$--${\rm R}_-$ and ${\rm R}_-$--${\rm NS}_-$ sectors. This spectrum is non-chiral after a dimensional reduction to $d=6$ and is therefore free of gravitational anomalies.

When expressed in terms of theta functions, the partition function assumes the simple expression
\begin{equation*}
\mathscr{Z}= -\, \frac{1}{4\, \tau_2^2} \frac{1}{(\eta \,\bar \eta )^{12}} \left[ \vartheta_3^4\,\bar \vartheta_4^4 - \vartheta_4^4\, \bar \vartheta_3^4 \right]^2\,.
\end{equation*}
Since the combinations
\begin{equation*}
\frac{\vartheta_3^4 (0|q) }{\eta ^{6} (q)} =q^{-1/4} \sum_{n=0}^\infty d_n \, q^{n/2}\,,
\qquad
\frac{\vartheta_4^4 (0|q) }{\eta ^{6} (q)} =q^{-1/4} \sum_{n=0}^\infty (-1)^n\, d_n \, q^{n/2}\,,
\end{equation*}
have Taylor expansions with identical coefficients modulo a sign, and in general
\begin{equation*}
\left(\sum_{n=0}^\infty \alpha_n \, x^n \right) \, \left( \sum_{n=0}^\infty \beta_n \, x^n \right) = \sum_{n=0}^\infty \gamma_n \, x^n\qquad
{\rm with}\quad
\gamma_n = \sum_{p=0}^n \alpha_p\, \beta_{n-p}\,,
\end{equation*}
one gets
\begin{equation*}
\mathscr{Z} = \,\frac{1}{4 \, \tau_2^2}\, \sum_{n,m=0}^\infty \left[  2 \, c_n \, c_m - \left( (-1)^m +(-1)^n \right) b_n \, b_m  
 \right] \, q^{(n-1)/2}\, \bar q^{(m-1)/2}  \,,
\end{equation*}
where
\begin{equation*}
b_n = \sum_{p=0}^n d_{p}\, d_{n-p}\quad {\rm and}
\qquad
c_n = \sum_{p=0}^n (-1)^p\, d_{p}\, d_{n-p}\,.
\end{equation*}
The constant term in the Fourier expansion can be readily computed and reads
\begin{equation*}
a_0 (\tau_2 ) = \frac{1}{2\,\tau_2^2}\, \sum_{m=0}^\infty \left( c_m^2 - (-1)^m\, b_m^2 \right)\, e^{-2\pi( m-1) \tau_2} \,.
\end{equation*}
Notice that  $b_0^2 = c_0^2$ is consistent with the fact that there are no physical tachyons in the spectrum, while 
$\frac{1}{2} (c_1^2 - b_1^2) = 2 d_0^2 \, d_1^2 =128$ correctly counts the graded number of massless degrees of freedom.

\medskip

\begin{figure}[h] 
	\begin{center}
		\includegraphics[scale=.9]{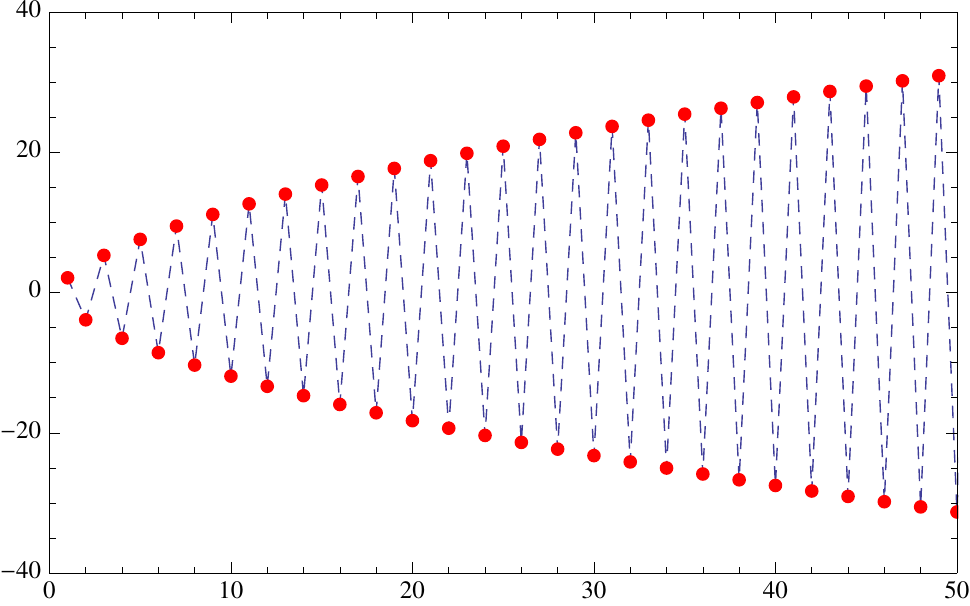}
		\caption{\small Logarithmic plot of the density of states ($\log_{10} | c^2_m - (-1)^m \, b_m^2 |$) for the Scherk-Schwarz reduction. To emphasize the alternating behaviour of the spectrum bosons (fermions) are assigned a positive (negative) sign.
		 		} 
	\end{center}
\end{figure}
 
The coefficients $c_m^2 - (-1)^m\, b_m^2$ are plotted in figure 4, and show that bosons and fermions alternate in this non-tachyonic vacuum, as expected from the {\it misaligned supersymmetry} or, alternatively, as a consequence of the oscillating behaviour of the dominant error term in equation (\ref{main}).

A numerical estimate of the integral of $\mathscr{Z} (\tau_1 , \tau_2 )$ over the fundamental domain yields
\begin{equation*}
\varOmega = \int_\mathscr{F} d\mu \, \mathscr{Z} (\tau_1 , \tau_2 ) \simeq 50\,,
\end{equation*}
while, from eq. (\ref{main}) and assuming true the Riemann hypothesis, $\rho_m = \frac{1}{2} +i \gamma_m$, we can write
\begin{equation*}
\begin{split}
g (\tau_2 ) &\equiv  \sum_{m=0}^\infty \frac{c_m^2 - (-1)^m\, b_m^2 }{2}\, e^{-2\pi( m-1) \tau_2} 
\\
&\sim \frac{3}{\pi} \, \varOmega\,\tau_2^2 + \tau_2^{11/4} \sum_{m=1}^\infty C_m \cos \left( \tfrac{1}{2} \, \gamma_m \, \log \, \tau_2 +\phi_m \right)\,.
\end{split}
\end{equation*}
It is clear from figure 5 that $g (\tau_2 )$ gives an excellent approximation of $\varOmega$ as $\tau_2 \to 0$. The oscillations around the value of the vacuum energy clearly indicates the phenomenon of {\it misaligned supersymmetry} and the fact that the dominant contribution to the error $(g(\tau_2 ) - \frac{3}{\pi}\, \varOmega \, \tau_2^2 )\, \tau_2^{-11/4} $ is related to the position of the non trivial zeroes of the Riemann $\zeta$- function.

In all the examples we have studied the results are constistent with the Riemann hypothesis. One may dare to hope that if the hypothesis is wrong a counter example will arise from such string theory considerations.

\medskip

\begin{figure}[h] 
	\begin{center}
		\includegraphics[scale=.65]{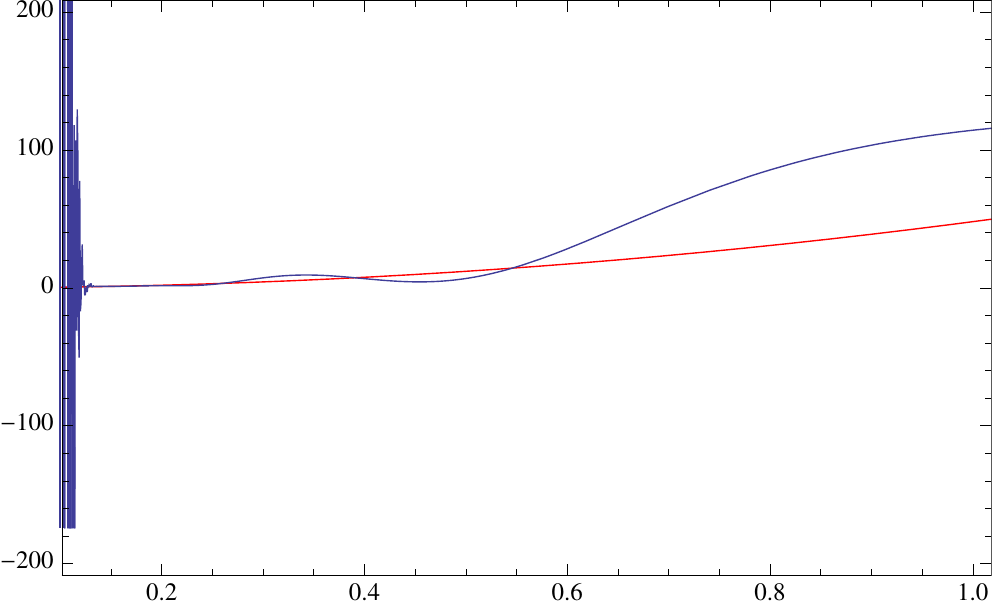} \hskip 20pt
		\includegraphics[scale=.77]{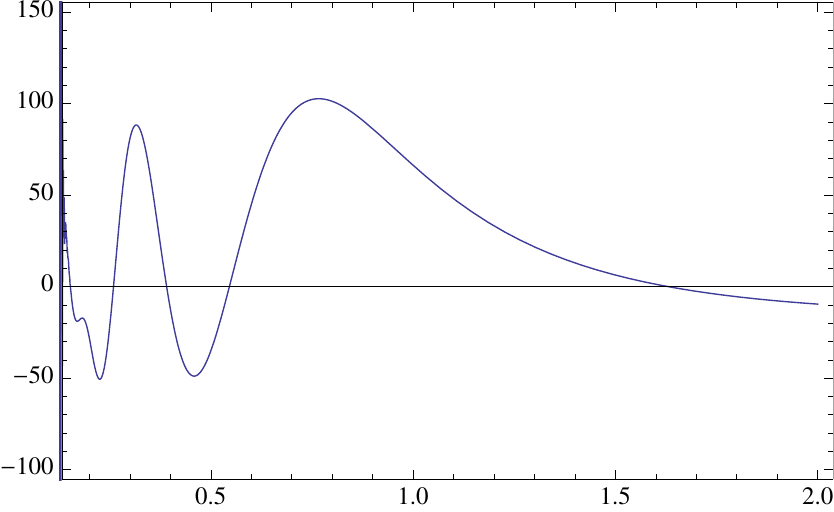} 
		\caption{\small The vacuum energy and its approximation for the Scherk-Schwarz dimensional reduction on the SO(8) lattice. In the left figure, the blue (red) line shows the behaviour of the function $g (\tau_2 )$ ($3 \,\tau^2_2 \, \varOmega /\pi$) for small values of $\tau_2$. The wild oscillations near $\tau_2 \simeq 0$ are a consequence of the truncation of the infinite sum in $g$. The right figure represents the error term $(g (\tau_2 ) - \frac{3}{\pi}\,\varOmega\,\tau_2^2 )\,\tau_2^{-11/4}$. The detailed structure of the oscillations results from the structure of the non-trivial zeroes of the Riemann $\zeta$-function, and is consistent with the Riemann hypothesis.
		 		} 
	\end{center}
\end{figure}

\section{A comment on functions of rapid growth at the cusp and on vacua with physical and unphysical tachyons}
\label{tachyons}

As already emphasised, the Rankin-Selberg-Zagier formula applies only to the class of automorphic functions that are at most of moderate growth at the cusp. In a string-theory language this translates into configurations where neither physical nor unphysical tachyons are present. 
However, although non-tachyonic vacuum configurations represent a vast class of models where stability is under control, at least classically, and several conclusions can be drawn about quantities of physical interest, these represent a small corner of string-theory vacua. In particular, heterotic strings {\it always} have unphysical tachyons, simply because the left-handed and right-handed vacuum states have different energies. In supersymmetric vacua they are harmless since their coefficients in the partition function are identically vanishing (as is the identically vanishing vacuum energy itself). However, when supersymmetry is broken, spontaneously or explicitly, they are generically present, and therefore invalidate the analysis presented in the previous sections.

Still, can one extend in some way the Rankin-Selberg-Zagier method to functions of rapid growth at the cusp?

Unfortunately, mathematical theorems are not available at present, although it is reasonable to assume that actually this is the case, at least for vacua with unphysical tachyons only, as we are going to show momentarily. 

Let us comment first on the configurations when physical tachyons are present in the spectrum. In this case it is hard to establish a direct link between the vacuum energy and the graded density of states, since in general, these are two divergent quantities. Indeed, the presence of tachyonic states signals the onset of some instability, and thus the perturbative description fails to apply. Nevertheless, it would be interesting to associate the existence or not of such an instability to UV properties of the spectrum and not only to the IR divergence of the free energy,
although a direct link between the IR divergence and the distribution of elementary degrees of freedom is difficult to establish. Let us consider, in fact, those configurations where supersymmetry is spontaneously broken (by finite temperature effects or by non-trivial boundary conditions), or more generally those non-supersymmetric tachyonic string vacua that can be obtained as freely acting orbifolds of some parent supersymmetric configurations. This is actually the case of most vacua of physical interest. In these cases, the free energy is divergent because of the tachyon, while the graded number of degrees of freedom is actually vanishing in the limit $\tau_2 \to 0$ since the new non-supersymmetric spectrum is obtained as an adiabatic perturbation of a supersymmetric one. As a result we expect that the two spectra enjoy the same {\it asymptotic supersymmetry}. Therefore, if a signature of the instability should be encoded in the distribution of the degrees of freedom it would be quite difficult to decipher, and a better understanding of the fine structure of the string spectrum is necessary.

Quite different is the case of vacua with unphysical tachyons. In fact they do not bring-in any evident instability to the string vacuum and hence we expect that the one-loop amplitudes be meaningful and somehow related to the graded density of states. This would be the case if, for example, automorphic functions describing theories with unphysical tachyons split as
\begin{equation}
f (\tau_1 , \tau_2) = f_{\rm singular} (\tau_1 , \tau_2 ) + f_{\rm moderate} (\tau_1 , \tau_2 ) \,, \label{splitting}
\end{equation}
where $f_{\rm moderate} (\tau_1 , \tau_2 )$ and $f_{\rm singular} (\tau_1 , \tau_2 )$ are each an automorphic function, the first at most of moderate growth at the cusp while the latter has vanishing constant term in the Fourier expansion
\begin{equation*}
\int_{-1/2}^{1/2} d\tau_1 \, f_{\rm singular} (\tau_1 , \tau_2 ) \equiv 0\,,
\end{equation*}
and is of rapid growth at the cusp.
In such a case, one could apply the Rankin-Selberg-Zagier method to $f_{\rm moderate} (\tau_1 , \tau_2 ) $ and establish the relation
\begin{equation}
\lim_{\tau_2 \to 0} \, g(\tau_2 ) = \frac{3}{\pi}\, \int_{\mathscr{F}} d\mu\, f_{\rm moderate} (\tau_1 , \tau_2 )\,,
\label{modifiedRSZ}
\end{equation}
where, as usual $g(\tau_2 ) = \int_{-1/2}^{1/2} d\tau_1 \, f (\tau_1 , \tau_2 )$. An example is the $j$ cusp form
\begin{equation*}
\begin{split}
j (\tau ) &= q^{-1} + \sum_{n=1}^\infty a_n \, q^n
\\
&= q^{-1} + 196\, 884\, q + 21\, 493\, 760\, q^2 + 864\,299\,970\, q^3 + \ldots\,,
\end{split}
\end{equation*}
that indeed grows exponentially as $\tau \to i\infty$, and has vanishing constant term. In this case, $f_{\rm moderate} \equiv 0$ while $f_{\rm singular} (\tau_1 , \tau_2 ) = j (\tau )$ and
\begin{equation*}
0= \lim_{\tau_2 \to 0}\, \int_{-1/2}^{1/2} d\tau_1 \, f (\tau_1 , \tau_2 ) = 
\lim_{\tau_2 \to 0}\, \int_{-1/2}^{1/2} d\tau_1 \, j(\tau )
= \frac{3}{\pi} \int_{\mathscr{F}} d\mu \, f_{\rm moderate} (\tau_1 , \tau_2 ) = 0\,.
\end{equation*}

Unfortunately, it is not known whether the splitting (\ref{splitting}) is natural or not in the class of modular forms of rapid growth, but in the following we will present a couple of string-theory examples where eq.  (\ref{splitting}) seems to occur and the Rankin-Selberg-Zagier method could be used to connect UV and IR properties of the partition function.

\subsection{The heterotic ${\rm O} (16) \times {\rm O } (16) $ non-supersymmetric vacuum}

As a first string-theory example where the splitting (\ref{splitting}) naturally occurs and the relation (\ref{modifiedRSZ}) is at work, let us consider the ${\rm O} (16) \times {\rm O } (16) $ heterotic string compactified on an eight-dimensional lattice that, for simplicity, we take at the SO(16) symmetry enhancement point. 

Actually, the  ${\rm O} (16) \times {\rm O } (16) $ heterotic string is the prototype example of a non-supersymmetric vacuum configuration, whose only tachyonic excitations do not respect level-matching and thus are not part of the physical spectrum. The model \cite{AlvarezGaume:1986jb}  involves a generalised GSO projection that results from a non-geometric ${\mathbb Z}_2$ orbifold of the supersymmetric ${\rm E}_8 \times {\rm E}_8$ theory, the generator being $(-1)^{F+\sum_{i=1,2} F_i}$ where $F$ is the space-time fermion number and $F_i$ is a fermion number acting on the internal degrees of freedom of the $i$-th ${\rm E}_8 \sim {\rm Spin} (16) /{\mathbb Z}_2$ group factor.
In terms of the level-one ${\rm SO} (2n)$ characters \cite{Angelantonj:2002ct}, the one-loop partition function reads
\begin{equation*}
\begin{split}
{\mathscr Z} =& \tau_2^{-4}\, (\eta \, \bar \eta )^{-8}\, 
\Bigl[ V_8 \, ( \bar O_{16}\, \bar O _{16} +\bar S_{16}\,\bar S_{16} ) - S_8\, (\bar O _{16} \, \bar S_{16} + \bar S_{16}\, \bar O _{16} )
\\
&+ O_8 \, (\bar V_{16} \,\bar C_{16} + \bar C_{16}\, \bar V _{16} ) - C_8 \, ( \bar V_{16} \,\bar V _{16} + \bar C_{16}\,\bar C_{16} ) \Bigr]\,,
\end{split}
\end{equation*}
from which it is straightforward to extract the massless spectrum comprising a ten-dimensio\-nal graviton, a dilaton, a Kalb-Ramond antisymmetric tensor, gauge bosons in the adjoint of the gauge group ${\rm O} (16) \times {\rm O } (16) $, left-handed fermions in the spinorial representations $(128,1)+(1,128)$ together with right-handed fermions in the bi-fundamental representation $(16,16)$. The spectrum is free of irreducible gauge and gravitational anomalies as a consequence of modular invariance, while the reduced anomalies are disposed of by a Green-Schwarz mechanism. 

After compactification to $d=2$ on the SO(16) lattice, the partition function reads
\begin{equation*}
\begin{split}
{\mathscr Z} =& 
\Bigl[ V_8 \, ( \bar O_{16}\, \bar O _{16} +\bar S_{16}\,\bar S_{16} ) - S_8\, (\bar O _{16} \, \bar S_{16} + \bar S_{16}\, \bar O _{16} )
+ O_8 \, (\bar V_{16} \,\bar C_{16} + \bar C_{16}\, \bar V _{16} ) 
\\
&- C_8 \, ( \bar V_{16} \,\bar V _{16} + \bar C_{16}\,\bar C_{16} ) \Bigr]
 \,
\left( |O_{16} |^2 + |V_{16}|^2 + |S_{16}|^2 + |C_{16}|^2 \right)
\\
=& \frac{1}{4}\, \frac{1}{\eta^{12}\,\bar \eta^{24}} \left[ \vartheta_3^4 \, (\bar\vartheta_3^8 - \bar\vartheta_2^8 ) \, \bar\vartheta_4^8 - \vartheta_4^4 \, (\bar\vartheta_4^8 - \bar\vartheta_2^8 ) \, \bar\vartheta_3^8 \right] \, \left(
|\vartheta_3 |^{16} + |\vartheta_4|^{16} + |\vartheta_2 |^{16} \right)
\,.
\end{split}
\end{equation*}
The advantage of studying a two-dimensional vacuum configuration resides in the fact that ${\mathscr Z}$ does not depend explicitly on $\tau_2$ and, therefore, it is relatively easy to identify $f_{\rm singular}$ and $f_{\rm moderate}$. In fact, the Taylor expansion 
\begin{equation*}
{\mathscr Z} = \frac{8}{\bar q} -1216 + 526\, 496 \, \bar q + \frac{4096\, \sqrt{ \bar q}}{\sqrt{q}} + 122\, 880 \,\sqrt{q\,\bar q}
+106\, 496 \, q + \frac{1024\, q}{\bar q} - 66\, 826\, 240\, q \, \bar q+\ldots \,,
\end{equation*}
suggests that $\bar q^{-1}$ is the leading (singular) term in $f_{\rm singular}$
whose unique (anti-holomorphic) modular completion with vanishing constant term is given by the familiar $j$ cusp form\footnote{Notice that here we are requiring that $f_{\rm singular}$ be a holomorphic automorphic cusp form. If we allow for non holomorphic functions the modular completion is not unique any longer. In any case, the choice of the decomposition (\ref{splitting}) does not matter.} \cite{serre}. It is then straightforward to write
\begin{equation*}
f_{\rm singular} ( \bar \tau ) =8\, j (\bar q )\,, \qquad
f_{\rm moderate} (\tau_1 , \tau_2 ) = {\mathscr Z} -8\, j ( \bar q )\,,
\end{equation*}
and relate the function
\begin{equation*}
g (\tau_2 ) =\int_{-1/2}^{1/2} d\tau_1 \, {\mathscr Z} (\tau_1 , \tau_2 )
\end{equation*}
to the modular integral of $f_{\rm moderate}$. In this case, we do not expect {\it asymptotic supersymmetry} since in two dimensions
\begin{equation}
\lim_{\tau_2 \to 0} g (\tau_2 ) = \frac{3}{\pi}\, \varOmega_p \equiv
\frac{3}{\pi} \int_{\mathscr F} d\mu\, \left( {\mathscr  Z} (\tau_1 , \tau_2 ) -8 \,j (\bar q ) \right) \simeq -1261
\not= 0\,. 
\label{hetRSZ}
\end{equation}
A rough numerical analysis of $g (\tau_2 )$ is plotted in figure 6. Unfortunately, due to the slow convergence of the relation 
(\ref{hetRSZ}) and to the limited computational power at our disposal, we cannot efficiently probe the region $\tau_2 \to 0$. 
The amplitude of the oscillations decrease very slowly, at most as $\tau_2^{3/4} $ if the Riemann hypothesis is true, while their frequencies are as usual given by the imaginary part of the zeroes of $\zeta^* (s)$.

\medskip

\begin{figure}[h] 
	\begin{center}
		\includegraphics[scale=.9]{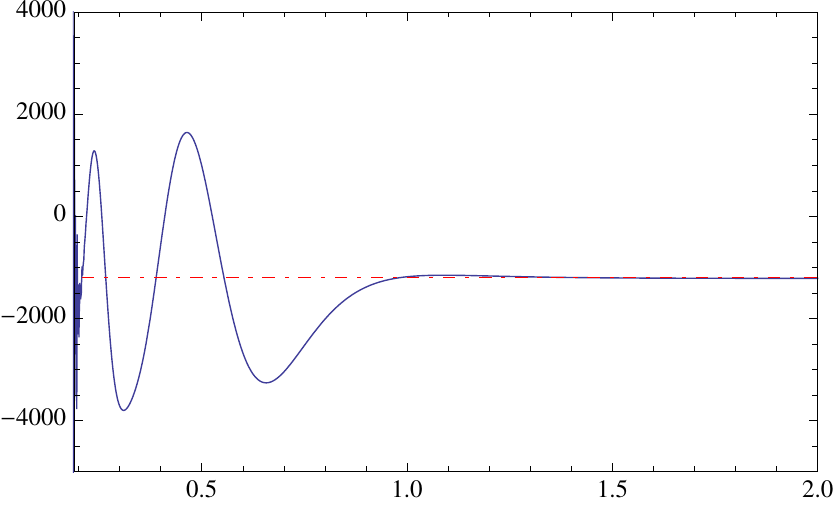}  
		\caption{\small The vacuum energy and its approximation for the non-tachyonic ${\rm O} (16) \times {\rm O} (16)$ heterotic string compactified on an eight-dimensional lattice at the SO(16) symmetry enhancement point. The figure represents the function $g (\tau_2 )$ (blue solid line) and the constant value of the {\it physical} vacuum energy density $\frac{3}{\pi}\,\varOmega_p$ (red dot-dashed line). The detailed structure of the oscillations results from the structure of the non-trivial zeroes of the Riemann $\zeta$-function, and is consistent with the Riemann hypothesis.
		 		} 
	\end{center}
\end{figure}

However, from figure 6 one can learn that the contribution to the {\it physical} free energy $\varOmega_p $ comes  almost entirely from the overall $-1216$ massless states (the only ones surviving the large-$\tau_2$ region), that yield $\varOmega_p^{\rm massless} = -1216\cdot \pi/3 \simeq -1320$, and thus contribute approximately $96\%$ to $\varOmega_p$. This result actually suggests that very massive states contribute very little to the integral over the fundamental domain. Therefore, the exact modular completion of the unphysical tachyon, while of crucial importance for the mathematical consistency of the conjecture (\ref{modifiedRSZ}) is to all practical purposes irrelevant. Actually, this example is just one representative in a class of two dimensional heterotic vacua without physical tachyons. In this class of vacua, $f_{\rm singular}$ is given by $d_{-1} \, j (\bar q )$, where $d_{-1}$ counts the number of unphysical tachyons, and their analysis follows step-by-step this $d=2$ ${\rm O} (16) \times {\rm O} (16) $ heterotic string example.

We can then consider the ten-dimensional ${\rm O} (16) \times {\rm O} (16) $ non-tachyonic heterotic string and Taylor expand the partition function as
\begin{equation}
\begin{split}
{\mathscr Z} =& \tau_2^{-4}\, (\eta \, \bar \eta )^{-8}\, 
\Bigl[ V_8 \, ( \bar O_{16}\, \bar O _{16} +\bar S_{16}\,\bar S_{16} ) - S_8\, (\bar O _{16} \, \bar S_{16} + \bar S_{16}\, \bar O _{16} )
\\
&+ O_8 \, (\bar V_{16} \,\bar C_{16} + \bar C_{16}\, \bar V _{16} ) - C_8 \, ( \bar V_{16} \,\bar V _{16} + \bar C_{16}\,\bar C_{16} ) \Bigr]
\\
=& \frac{1}{2\,\tau_2^4}\, \frac{1}{\eta^{12}\,\bar \eta^{24}} \left[ \vartheta_3^4 \, (\bar\vartheta_3^8 - \bar\vartheta_2^8 ) \, \bar\vartheta_4^8 - \vartheta_4^4 \, (\bar\vartheta_4^8 - \bar\vartheta_2^8 ) \, \bar\vartheta_3^8 \right]\,,
\\
=& \frac{1}{\tau_2^4}\, \left[ \bar q ^{-1}\, \sum_{n=0}^\infty d_{2n} \, q^n -
\sum_{m=0}^\infty \sum_{n=-1}^\infty \frac{ (-1)^m + (-1)^n}{2} \, d_n \, c_m \, q^{n/2}\, \bar q ^{m/2} \right]  \,.
\end{split}
\label{heterotic}
\end{equation}
$c_n$ and $d_n$ are the coefficients in the Taylor expansions of
\begin{equation*}
\begin{split}
\frac{\vartheta_3^4}{\eta^{12}} &= \sum_{n=-1}^\infty d_n \, q^{n/2}\,, 
\\
\frac{\vartheta_4^4}{\eta^{12}} &= \sum_{n=-1}^\infty (-1)^{n-1}\, d_n \, q^{n/2}\,, 
\end{split}
\qquad
\begin{split}
\frac{(\bar \vartheta_2^8 -\bar\vartheta_4^8 )\, \bar\vartheta_3^8}{\bar \eta^{24}} &= -\bar q ^{-1}+ \sum_{n=0}^\infty  c_n \, \bar q^{n/2}\,,
\\
\frac{(\bar \vartheta_3^8 -\bar\vartheta_2^8 )\, \bar\vartheta_4^8}{\bar \eta^{24}} &= \bar q ^{-1}+ \sum_{n=0}^\infty  (-1)^{n-1}\, c_n \, \bar q^{n/2}\,.
\end{split}
\end{equation*}

To apply the conjecture (\ref{modifiedRSZ}) one should at this point identify the two automorphic functions $f_{\rm moderate}$ and $f_{\rm singular}$. While it is obvious that $f_{\rm singular}$ starts with the unphysical tachyon $\tau_2^{-4}\, \bar q ^{-1}$, it is not straightforward to determine its modular completion, because of the direct dependence on $\tau_2$. However, as we have observed in the previous example, assuming the splitting (\ref{splitting}) does occur also in this case,
the correct identification of the Taylor expansion of $f_{\rm singular}$ is not important at the practical level, 
since only the light states give a sizeable contribution to $\varOmega_p$. 

\medskip

\begin{figure}[h] 
	\begin{center}
		\includegraphics[scale=.9]{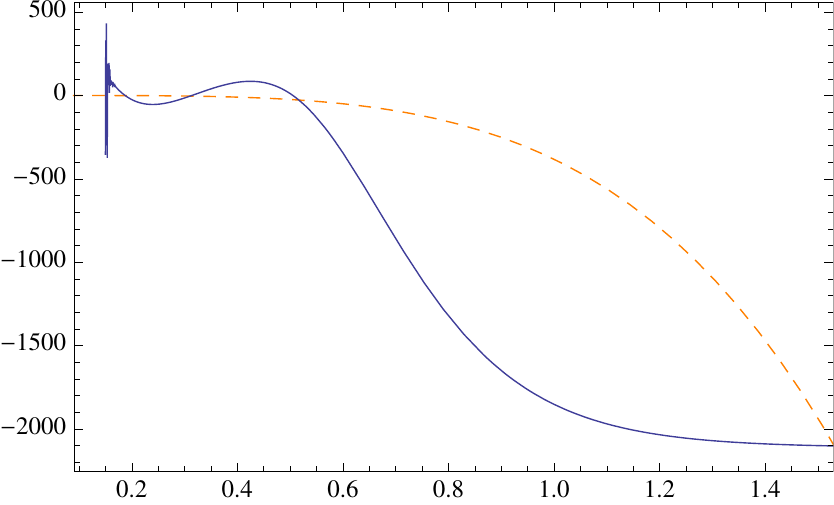} 
		\caption{\small The vacuum energy and its approximation for the non-tachyonic ${\rm O} (16) \times {\rm O} (16)$ heterotic string. The blue solid line shows the behaviour of the function $g (\tau_2 )$  for small values of $\tau_2$, while the orange dashed  line shows the function $3 \,\tau^4_2 \, \varOmega_p /\pi$.
The wild oscillations near $\tau_2 \simeq 0$ are a consequence of the truncation of the infinite sum in $g$. The detailed structure of the oscillations results from the structure of the non-trivial zeroes of the Riemann $\zeta$-function, and is consistent with the Riemann hypothesis.
		 		} 
	\end{center}
\end{figure}

Omitting the unphysical tachyon from ${\mathscr Z}$, one can numerically approximate the {\it physical} vacuum energy as
\begin{equation*}
\varOmega_p = \int_{\mathscr F} d\mu \, \left[ {\mathscr Z} (\tau_1 , \tau_2 ) - \frac{d_0}{\tau_2^4\, \bar q}\right] \simeq -498 \,.
\end{equation*}
For the constant term in the Fourier expansion of ${\mathscr Z}$ there is no ambiguity
\begin{equation*}
g (\tau_2 ) =  \sum_{m=0}^\infty (-1)^{m-1}\, d_m \, c_m \, e^{-2\pi m \tau_2}\,,
\end{equation*}
since $f_{\rm singular}$ is assumed to be a cusp form. Therefore, following the previous observations, it would be tempting to apply the Rankin-Selberg-Zagier method also to this example and write
\begin{equation} 
g (\tau_2 ) \sim \frac{3}{\pi}\, \varOmega_{p} \, \tau_2^4 + {\rm error} \,,
\label{hetrelation}
\end{equation}
as $\tau_2 \to 0$, where the error term would be of order $o ( \tau_2^{19/4} )$ if eq. (\ref{main}) holds and the Riemann hypothesis is correct. 
The validity of the relation (\ref{hetrelation}) is again supported by an independent evaluation of $g (\tau_2 )$, as can be seen from figure 7. Unfortunately the available accuracy of the numerical evaluation is not terrific. Nevertheless, the behaviour in figure 7 encourages further explorations  both from the mathematical side and from the numerical side in order to establish a clean UV-IR relation  in the class of vacua with unphysical tachyons.

\vskip 24pt


\section*{Acknowledgements}

We thank Jose Barbon, Sergio Cacciatori, David Kazhdan, Peter Sarnak, Nathan Seiberg, Stephen Shenker, and Gabriele Veneziano for enlightening discussions. M.C. thanks Don Zagier for pointing out to him his results on the dependence of horocycle averages of automorphic functions on the non trivial zeros of the Riemann zeta function. C.A. would like to thank the Racah Institute of Physics of the Hebrew University of Jerusalem, the Theory Unit at CERN and  the KITP, Santa Barbara, and the organisers of the workshop ``Strings at the LHC and in the Early Universe'' for their warm hospitality during various stages of this collaboration. The work of CA is supported in part by the Italian MIUR-PRIN contract 20075ATT78 and in part by the ERC Advanced Grant no. 226455, ``Supersymmetry, Quantum Gravity and Gauge Fields'' (SUPERFIELDS). M.C. would like to thank the Racah Institute of Physics of the Hebrew University of Jerusalem for support, and the ESI Schr\"odinger Center for Mathematical Physics in Vienna and the Theory Unit at CERN for hospitality and support during the final stage of this work. The work of M.C. was supported during different stages by the Italian MIUR-PRIN contract 20075ATT78 at the University of Milano Bicocca, by the Theory Unit at CERN and  by Superstring Marie Curie Training Network under the contract MRTN-CT-2004-512194. M.C. is presently supported by a ''Angelo Della Riccia'' fellowship at the University
of Amsterdam. The work of S.E. was partially supported by the Israel Science Foundation, the Einstein Center in the Hebrew University, and by a grant of DIP (H.52). The work of E.R. was partially supported by the European Union Marie Curie RTN network under contract MRTN-CT-2004-512194, the American- Israel Bi-National Science Foundation, the Israel Science Foundation, The Einstein Center in the Hebrew University, and by a grant of DIP (H.52).

\end{document}